\documentclass[secnumarabic,amssymb,nobibnotes,aps,prl,preprint,letterpaper,preprintnumbers,amsmath,superscriptaddress,floatfix,notitlepage]{revtex4-1}

\usepackage{wrapfig}
\usepackage{amsmath}
\usepackage{enumitem}
\usepackage{multirow}
\usepackage[letterpaper,margin={1in,1in}]{geometry}
\usepackage{graphicx}
\usepackage{epstopdf}
\usepackage{lineno}
\usepackage{appendix}
\usepackage{longtable}
\usepackage{array}
\usepackage{siunitx}
\usepackage[dvipsnames]{xcolor}
\usepackage{lineno}
\usepackage{isotope}
\setcitestyle{super}
\usepackage{multirow}
\DeclareSIUnit\mrad{\milli\rad}

\usepackage{soul}
\usepackage{notes2bib}

\begin{document}

\setlength{\LTcapwidth}{6.5in}

\title{Vibrational fingerprints of ferroelectric HfO$_2$}

\affiliation{Department of Physics and Astronomy, University of Tennessee, Knoxville, Tennessee, 37996, USA}
\affiliation{Department of Physics and Astronomy, Rutgers University, Piscataway, New Jersey 08854, USA}
\affiliation{Rutgers Center for Emergent Materials, Rutgers University, Piscataway, New Jersey 08854, USA}
\affiliation{Department of Chemistry, University of Tennessee, Knoxville, Tennessee, 37996, USA}
\affiliation{These authors contributed equally to this work}

\author{Shiyu Fan}
\affiliation{Department of Physics and Astronomy, University of Tennessee, Knoxville, Tennessee, 37996, USA}
\affiliation{These authors contributed equally to this work}

\author{Sobhit Singh}
\affiliation{Department of Physics and Astronomy, Rutgers University, Piscataway, New Jersey 08854, USA}
\affiliation{These authors contributed equally to this work}

\author{Xianghan Xu}
\affiliation{Department of Physics and Astronomy, Rutgers University, Piscataway, New Jersey 08854, USA}
\affiliation{Rutgers Center for Emergent Materials, Rutgers University, Piscataway, New Jersey 08854, USA}

\author{Kiman Park}
\affiliation{Department of Chemistry, University of Tennessee, Knoxville, Tennessee, 37996, USA}

\author{Yubo Qi}
\affiliation{Department of Physics and Astronomy, Rutgers University, Piscataway, New Jersey 08854, USA}

\author{S. W. Cheong}
\affiliation{Department of Physics and Astronomy, Rutgers University, Piscataway, New Jersey 08854, USA}
\affiliation{Rutgers Center for Emergent Materials, Rutgers University, Piscataway, New Jersey 08854, USA}

\author{David Vanderbilt}
\affiliation{Department of Physics and Astronomy, Rutgers University, Piscataway, New Jersey 08854, USA}

\author{Karin M. Rabe}
\email{rabe@physics.rutgers.edu}
\affiliation{Department of Physics and Astronomy, Rutgers University, Piscataway, New Jersey 08854, USA}

\author{J. L. Musfeldt}
\email{musfeldt@utk.edu}
\affiliation{Department of Physics and Astronomy, University of Tennessee, Knoxville, Tennessee, 37996, USA}
\affiliation{Department of Chemistry, University of Tennessee, Knoxville, Tennessee, 37996, USA}

\begin{abstract}
{\bf 
Hafnia (HfO$_2$) is a promising material for emerging chip applications due to its high-$\kappa$ dielectric behaviour, 
suitability for  negative capacitance heterostructures, scalable ferroelectricity, and  silicon compatibility.
The lattice dynamics along with phononic properties such as thermal conductivity, contraction, and  heat capacity are under-explored, primarily due to the absence of high quality single crystals. 
Herein, we report the vibrational properties of a series of HfO$_2$ crystals stabilized with yttrium (chemical formula HfO$_2$:$x$Y, where $x$ = 20, 12, 11, 8, and 0\%) and  
compare our findings with a symmetry analysis and 
lattice dynamics calculations. 
We untangle the effects of Y by testing our calculations against the measured Raman and infrared spectra of the cubic, antipolar orthorhombic, and monoclinic phases and then proceed to reveal the signature modes of polar orthorhombic hafnia.
This work provides a spectroscopic fingerprint for several different phases of HfO$_2$ and paves the way for an analysis of mode contributions to high-$\kappa$ dielectric and ferroelectric properties for chip technologies. }

\end{abstract}

\maketitle

\section{Introduction}

Phonons in HfO$_2$ are key to understanding competing phases and physical properties of this emerging scientifically and technologically important material, unlocking the door to discoveries in fields as diverse as thin film dielectrics and nanoscale devices and photonics.~\cite{Wilk2001,Hoffmann_review2020,Clima2014,Boscke2011, Park2018, Zhou2015, Alessandri2018,Choe2021,Wei2018,Cheema2020,Park_aelm2019, Lee2020, Xu2021, Mikolajick2021, Jiang_review2021, Mandal2016, HuanPRB2014,Breyer2021} 
It has recently been discovered that flat phonon bands have a direct connection to the unique energy landscape in the vicinity of the ferroelectric phase.~\cite{Lee2020} 
This originates from the alternatively ordered spacer and polar HfO$_2$ layers, which creates very localized electric dipoles within the irreducible half-unit cell widths ($\approx$3 \AA) of HfO$_2$. \cite{Lee2020} 
The nanometer-scale dipoles are individually switchable without any net energy cost, though with energy barriers that correspond to very large coercive fields. \cite{Zhou2015,Alessandri2018, Choe2021}
Physical properties determined by phonons include heat capacity and thermal conductivity behavior, with
heat management in memory and logic devices being vital for technological applications. 

\begin{figure*}[tbh]
\centering
\includegraphics[width=5.2in]{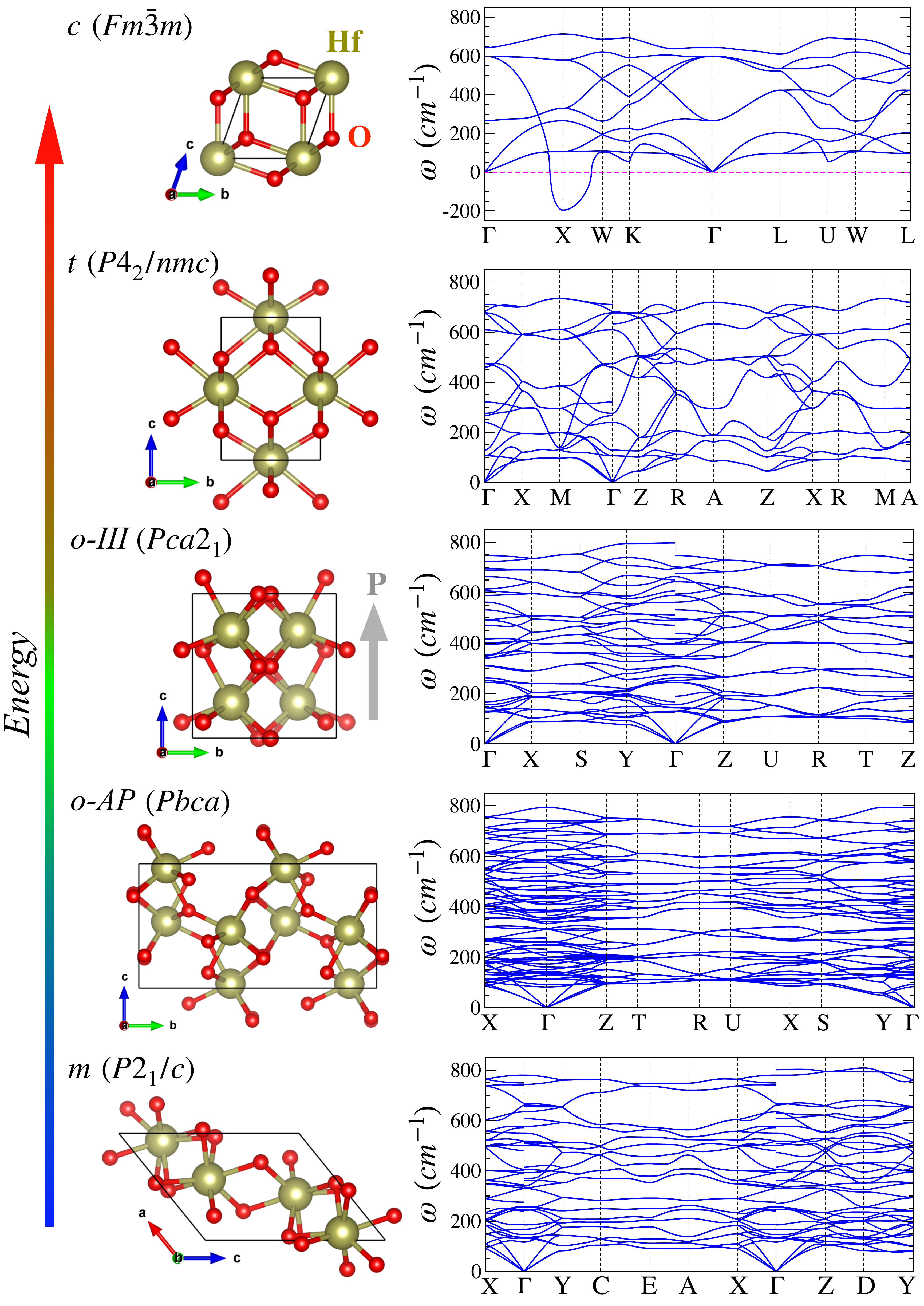}
\caption{ {\bf Crystal structure and calculated phonon dispersions of the five studied phases of bulk HfO$_2$.} These include:  cubic (c), tetragonal (t), orthorhombic polar (o-III), orthorhombic antipolar (o-AP), and monoclinic (m),  arranged in the order of their increasing formation energy. The c phase has the highest formation energy, whereas the m phase has the lowest formation energy. 
The gray arrow denotes the direction of the ferroelectric polarization (${\bf P}$) in the o-III phase.   }
\label{fig:struct}
\end{figure*}

HfO$_2$ is highly polymorphic, with many competing phases generated by different distortions of the high-symmetry cubic fluorite structure.~\cite{HuanPRB2014,park_AdvMat2018,QiPRL2020,QiPRB2020, Mandal2016, Park_aelm2019,Cojocaru2019,delodovici2021trilinear,Jiang_review2021,Xu2021, Mikolajick2021}  The cubic (c) Fm$\bar{3}$m fluorite phase itself is stabilized only at very high temperatures ($>$ 2900 K).~\cite{Wang_jms1992, HuanPRB2014,Park2018} 
Upon cooling, this phase transforms to a lower symmetry tetragonal (t) P4$_2$/nmc phase in temperature range 2900 -- 2073 K. 
At room temperature, hafnia stabilizes in a monoclinic (m) P2$_1$/c phase, which is the ground state phase of bulk hafnia at ambient conditions. Furthermore, two orthorhombic metastable polar (o-III) Pca2$_1$ and antipolar (o-AP) Pbca phases of hafnia are reported at higher pressures.~\cite{Arashi1992,JayaramanPRB1993,HuanPRB2014,Mandal2016,Jiang_review2021, Low2021} 
Figure~\ref{fig:struct} summarizes the crystal structures of these five different phases along with their calculated phonon dispersions, showing that all except the cubic phase are locally stable.
The cubic-tetragonal phase transition is driven by an unstable zone-boundary $X_2^{-}$ phonon mode of the cubic phase. 
Ferroelectricity in the polar o-III phase of hafnia can be understood in reference to a high-symmetry centrosymmetric tetragonal phase.
Remarkably, there is no unstable polar mode in the high-symmetry tetragonal reference structure. Rather, first principles calculations have shown that the key factor in stabilizing the orthorhombic polar phase is a strong trilinear coupling among polar,  
nonpolar, 
and antipolar 
phonon modes.~\cite{reyeslillo2014,delodovici2021trilinear}
We note that the presence of oxygen vacancies, doping, substrate-induced strain, and disorder have been reported to stabilize various other phases of hafnia,~\cite{Shimizu2016,park_AdvMat2018,Wei2018,Park_aelm2019,Zhang2021,Wang_perspective2021, QiPRL2020,QiPRB2020, RushchanskiiPRL2021, Mikolajick2021,Dutta2020,Choe2021} which are not within the scope of the present investigation.

%
For almost a decade, the observation of ferroelectricity in hafnia has been limited only to thin films.~\cite{Boscke2011, Zhou2015,Shimizu2016, Park_JMCC2017, Park2018, Wei2018, Zhang2021, RushchanskiiPRL2021, Fina_ACS2021, Wang_perspective2021}
Stabilization of the orthorhombic polar phase in bulk form is not trivial.~\cite{Xu2021, Mikolajick2021} This is because the orthorhombic polar phase is predicted to be metastable, and it requires very high temperatures to form.~\cite{HuanPRB2014} 
There have been a number of attempts to prepare the elusive ferroelectric phases of hafnia at ambient conditions. 
Ultra-thin films HfO$_2$:Si have been synthesized to obtain the ferroelectric phases,~\cite{Boscke2011,Zhou2015, Alessandri2018, Shimizu2016,Park_JMCC2017,Park2018,Cheema2020,Fina_ACS2021,Wang_perspective2021, Zhang2021} although it is challenging to characterize these phases due to their small domain size and substrate effects. 
A rhombohedral ferroelectric phase was recently reported in epitaxially strained Hf$_{0.5}$Zr$_{0.5}$O$_2$ thin films also illustrating the complex interplay between the film and substrate.~\cite{Wei2018,Zhang2021} First principles calculations show that distinct competing ferroelectric phases can be stabilized in hafnia via epitaxial strain. \cite{QiPRL2020} 
Ferroelectricity in hafnia depends intimately on the concentration of oxygen vacancies as well. \cite{Nukala2021, RushchanskiiPRL2021} 

Recently, Xu {\it et al.}~demonstrated the stabilization of the polar o-III phase of hafnia as well as the antipolar o-AP phase in bulk single crystals grown using laser-diode-heated floating zone techniques.~\cite{Xu2021}  
Yttrium substitution and a rapid cooling rate are  key to realizing the polar o-III and antipolar o-AP phases at ambient conditions.~\cite{Xu2021} 
Ferroelectricity in the o-III phase was experimentally demonstrated with a switchable polarization and a magnitude 3\,$\mu$C cm$^{-2}$ together with a coercive electric field of 4\,MV cm$^{-1}$.~\cite{Xu2021}
%
%
The availability of high-quality single crystals creates a number of exciting opportunities to examine the properties of this family of materials - especially those of the more elusive members.  
While phonons in the cubic, tetragonal, and monoclinic phases have been investigated in prior work, \cite{Anastassakis1975, Kourouklis1991,Arashi1992, CarlonePRB1992, JayaramanPRB1993, Kim1997, ZhaoDVPRB2002, Belo2012, Mandal2016, Kumar2015, Singhal2019, Cojocaru2019, Borowicz2013} the lattice dynamics of the orthorhombic polar and antipolar phases are wholly unexplored.
In addition to providing a spectroscopic fingerprint for the different phases of HfO$_2$, this work opens the possibility of analyzing structure-property trends in the phonon excitation spectrum and Born effective charges, as in a recent work on the heavy chalcogenide 1T-HfS$_2$ in which it was found that the enhanced Born charge of hafnium is attributable to polar displacement-induced charge transfer from sulfur to hafnium. \cite{Neal2021}

In this work, we build upon these recent advances by measuring the infrared- and Raman-active phonons in the two orthorhombic phases of hafnia. We compare our results with the spectroscopic response of the cubic and monoclinic phases as well as a symmetry analysis and complementary lattice dynamics calculations, including analysis of the anomalous Born effective charges.
Taken together, our work places the vibrational properties of HfO$_2$ on a secure footing and paves the way for an advanced understanding of properties that depend upon the fundamental excitations of the lattice.  


\section{Results and Discussion}

\subsection*{Understanding the spectroscopic properties of cubic HfO$_2$:20\%Y}

\begin{figure*}[tbh]
\centering
\includegraphics[width=6.5in]{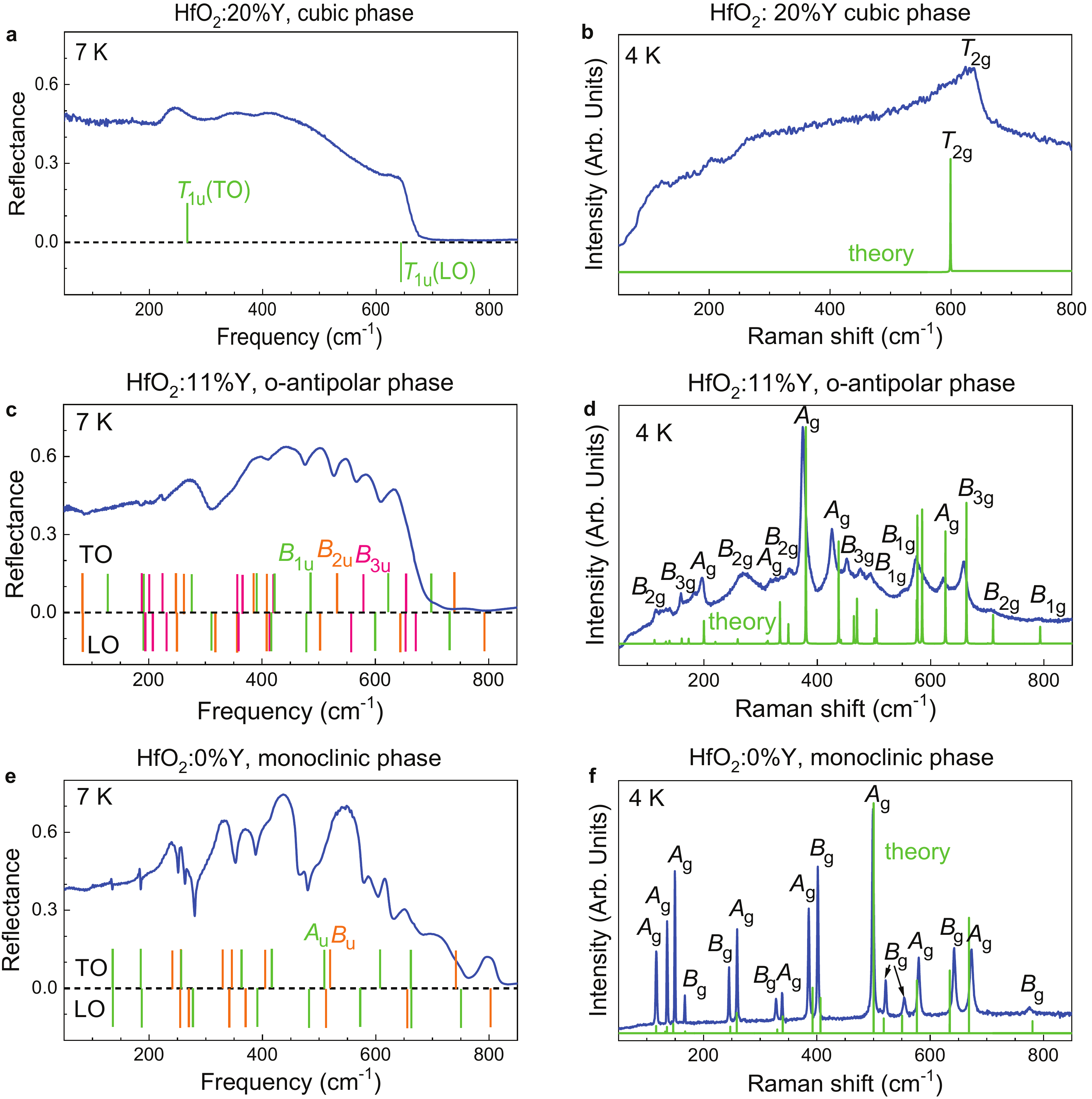}
\caption{ {\bf Infrared and Raman responses of the cubic, orthorhombic-antipolar, and monoclinic phases of HfO$_2$.} {\bf a,} Infrared reflectance spectrum of cubic HfO$_2$:20\%Y at 7 K. 
{\bf b,} Raman spectrum of cubic HfO$_2$:20\%Y  at 4 K. 
{\bf c,} Infrared reflectance spectrum of o-AP HfO$_2$:11\%Y  at 7 K. 
{\bf d,} Raman spectrum of o-AP HfO$_2$:11\%Y  at 4 K. 
{\bf e,} Infrared reflectance spectrum of monoclinic HfO$_2$  at 7 K. 
{\bf f,} Raman spectrum of monoclinic HfO$_2$  at 4 K. 
The calculated Raman response of pure HfO$_2$ 
is plotted along with the measured spectrum in green. 
The theoretical peak positions of the infrared TO and LO modes in pure HfO$_2$ are marked using vertical upward and downward lines, respectively.
Details are available  in the Supplementary Information. 
}
\label{HfO2 20Y and 11Y summary}
\end{figure*}

In order to investigate the lattice dynamics in different phases of hafnia, we systematically performed infrared reflectance and Raman-scattering measurements in near normal incidence and back-scattering geometry, respectively, using unpolarized light. 
Figure \ref{HfO2 20Y and 11Y summary}a,b displays the infrared reflectance and Raman scattering response of cubic HfO$_2$:20\%Y. 
Only two triply-degenerate vibrational modes are predicted at zone center in the pure cubic HfO$_2$ phase.
The T$_{1u}$ mode at 266 cm$^{-1}$ and T$_{2g}$ mode at 599 cm$^{-1}$ are infrared- and Raman-active, respectively [Supplementary Table 5]. However, they can both be active in the experimental infrared and Raman spectra because of local symmetry-breaking induced by yttrium incorporation [Supplementary Figure. 1]. The T$_{1u}$ mode splits into longitudinal optic (LO) and two transverse optic (TO) modes near the zone center at nonzero wavevector ($\bf {q \rightarrow 0}$). The calculated frequency of the T$_{1u}$(LO) mode is 643 cm$^{-1}$. In our measured infrared reflectance spectrum, we clearly observe the signatures of the T$_{1u}$(TO) and T$_{1u}$(LO) modes, even though the T$_{1u}$(LO) mode overlaps with the nearby T$_{2g}$ mode. The overlap between the two modes broadens the observed feature around 620 cm$^{-1}$ in the infrared reflectance spectrum. The T$_{2g}$ mode is also clearly detected in the measured Raman response. 

The overall agreement between our measurements and the predicted phonon frequencies is reasonable. 
But more than a simple match, these data reveal the challenges with Y incorporation - even in this  straightforward cubic system. 
The Raman scattering spectrum, for instance, has a noticeable fluorescence background that varies with excitation laser wavelength, and the scattering intensity depends upon the measurement spot due to surface roughness and cracking.  
At the same time, a  weak metallic response develops in the reflectance spectrum due to the Y inclusion. 
This type of weak metallicity is seen in other materials including Y-stabilized zirconia (ZrO$_2$:Y$_2$O$_3$) due to oxygen deficiencies~\cite{Liu1988,Peng2021,Frederick2008} as well as Nb-substituted EuTiO$_3$ and Pb-substituted BaPb$_x$Bi$_{1-x}$O$_3$ due to added carriers. \cite{Yokosuk2021,GERVAIS2002,PETIT1999} 
In any case, in our results, the over-damped Drude\cite{GERVAIS2002} partially screens the T$_{1u}$ phonon. 
At the same time, Y incorporation breaks local symmetry and activates several additional vibrational modes arising from Y sublattice. This is apparent in both the infrared and Raman response and occurs primarily below 400 cm$^{-1}$. 
This is consistent with our calculations for the ground state configuration of Y-substituted cubic HfO$_2$ [Supplementary Figure. 1]. 
Fortunately,  weak metallicity and the presence of a few additional vibrational features do not interfere with identifying the fundamental T$_{1u}$ and T$_{2g}$ vibrational modes in cubic HfO$_2$:20\%Y. 
These effects are diminished (but not absent) in the orthorhombic materials discussed below because the overall level of Y incorporation is significantly lower and therefore much less important.

\subsection*{Testing our predictions on antipolar orthorhombic HfO$_2$:11\%Y} 

Figure \ref{HfO2 20Y and 11Y summary}c displays the infrared spectrum of HfO$_2$ with 11\% Y substitution. 
This material is orthorhombic and antipolar (Pbca, o-AP). 
The weak metallic background in the infrared reflectance is overall reduced compared to that in HfO$_2$:20\%Y sample, and phonon screening is considerably weaker. 
As a result, all of the infrared-active phonons are sharper and better resolved. 
This effect is even more obvious in the 8\% sample [Supplementary Figure. 4], where the orthorhombic and monoclinic phases are mixed. 
The comparison between experiment and theory is good overall, even though the infrared-active phonons are partially screened by the weak metallic background induced by Y incorporation. 
Fortunately, the Raman-active phonons are well-resolved, providing an even better opportunity to compare  experiment and theory. 
Figure \ref{HfO2 20Y and 11Y summary}d displays the Raman scattering spectrum of HfO$_2$:11\%Y. 
Numerous Raman-active peaks are present due to the lower crystal symmetry compared to the cubic case, and the fluorescence background has been dramatically reduced using a longer wavelength laser ($\lambda$ = 532 nm). 
Compared to HfO$_2$:20\%Y, there are fewer phonon modes activated due to Y incorporation and the subsequent symmetry breaking. 
Overall, our experimental results nicely match the theoretical predictions 
in terms of both frequency and intensity.

\subsection*{Lattice dynamics of the monoclinic phase of HfO$_2$:0\%Y}  

The monoclinic phase is the ground state of bulk hafnia, so it is easy to stabilize this phase in single-crystal form without any Y substitution. 
Figure \ref{HfO2 20Y and 11Y summary}e,f summarizes the infrared reflectance and Raman scattering response of this phase. 
Since there is no Y incorporation, the infrared-active phonons are well defined and unscreened, and the fluorescence background observed in the samples containing 20 and 11\% Y is absent. 
The overall agreement between the measured and calculated pattern of excitations for pure hafnia is excellent and in perfect agreement with prior literature. \cite{Kumar2015,Singhal2019, ZhaoDVPRB2002}  
This illustrates the predictive power of theory and quality of vibrational spectra that should be available once the need for Y inclusion in the crystal growth process is eliminated. 

\subsection*{Revealing the vibrational properties of orthorhombic polar HfO$_2$:12\%Y} 

\begin{figure*}[tbh]
\centering
\includegraphics[width=6.5in]{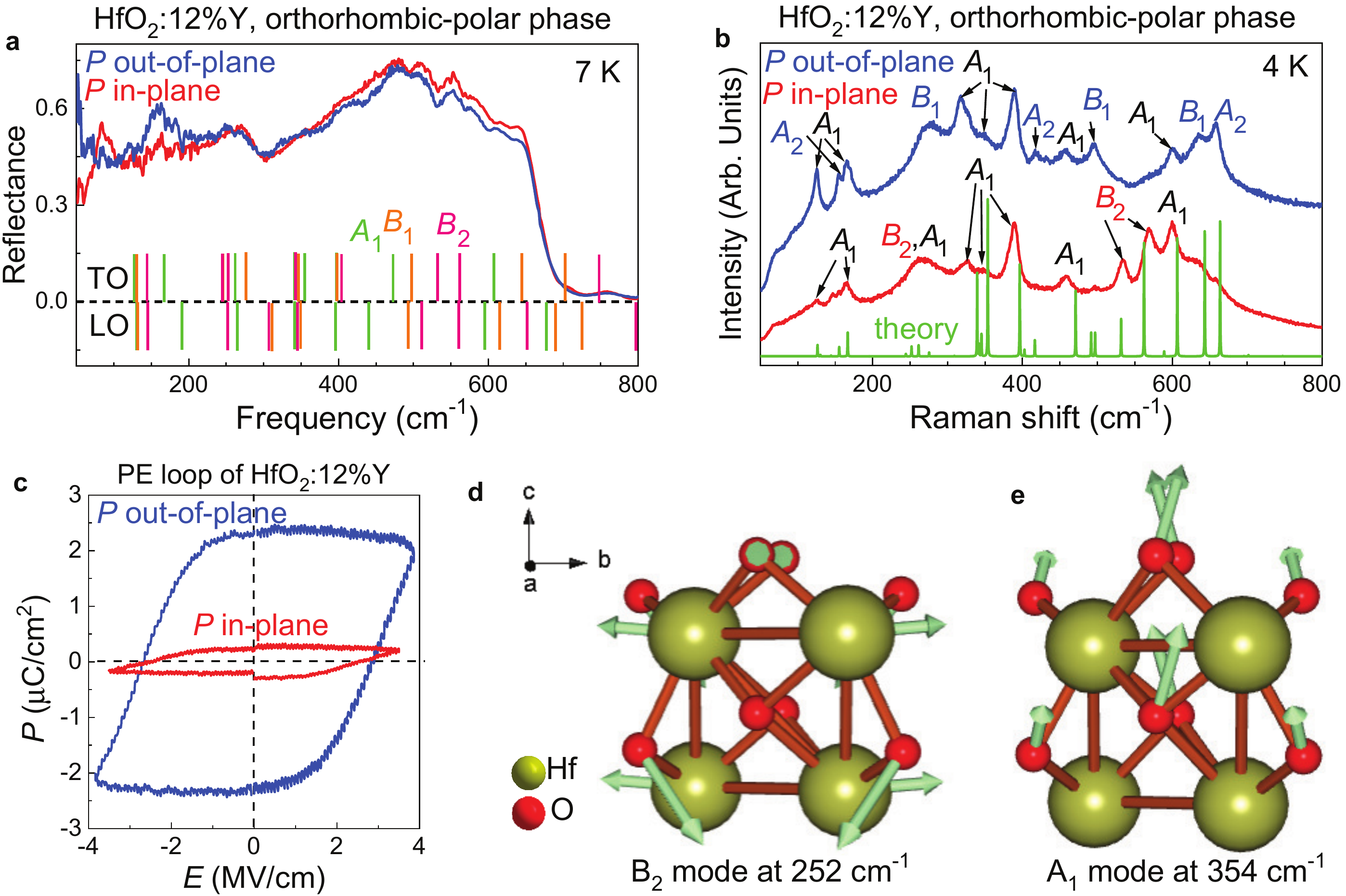}
\caption{ {\bf Infrared and Raman responses of the orthorhombic-polar HfO$_2$.} {\bf a,} 
Infrared reflectance spectra of HfO$_2$:12\% Y for in- and out-of-plane ferroelectric polarization samples measured at 7 K, compared with the theoretically calculated infrared TO and LO peak positions in the pure o-III phase. 
{\bf b,} Raman spectra of HfO$_2$:12\% for the in- and out-of-plane polarization samples measured at 4 K, compared with the theoretically calculated Raman spectrum of the pure o-III phase. The measurement geometries are summarized in Table~\ref{summary of HfO2} and Supplementary Figure. 3.  {\bf c,} Electric polarization hysteresis loops of the HfO$_2$:12\%Y. Red and blue colors indicate polarization along the out-of- and in-plane direction, respectively. {\bf d,e} Calculated phonon displacement patterns for the 252 cm$^{-1}$ Hf-Hf breathing mode, and the signature polar mode of the o-III phase at 354 cm$^{-1}$. Animations are available in the Supplementary Note 6.}
\label{HfO2 12Y summary}
\end{figure*}

After developing an in-depth understanding of the lattice dynamics and vibrational signatures of the above-studied phases of hafnia, we turn our attention to the polar orthorhombic (o-III) phase, which is central to this study.  
Figure~\ref{fig:struct} displays the crystal structure of the o-III phase, which was experimentally stabilized by 12\% Y substitution on the Hf lattice at rapid cooling conditions.~\cite{Xu2021}
A complex trilinear coupling involving a zone-center polar ($\Gamma_{3}^{-}$), 
and zone-boundary nonpolar ($Y_{2}^{+}$)
and antipolar ($Y_{4}^{-}$) 
phonon modes is responsible for the ferroelectric polarization in the o-III phase.~\cite{reyeslillo2014, delodovici2021trilinear} 
More simplistically, one can attribute ferroelectricity in this phase to the polar distortion of oxygen atoms along the $\vec{c}$ axis (in reference to a centrosymmetric tetragonal phase) as shown in Fig.~\ref{HfO2 12Y summary}(e), resulting in a spontaneous ferroelectric polarization ({\textbf{P}}) parallel to the $\vec{c}$ lattice vector (or $\hat{z}$ direction).~\cite{Lee2020, delodovici2021trilinear}

The primitive unit cell of o-III phase 
contains a total of 12 atoms (4 formula units) resulting in 33 optical phonon modes having following irreducible representations at the zone center: 
\begin{equation}
\Gamma\textsubscript{$o$-{\it III}} = 
8\,A_1 \oplus  9\,A_2 \oplus 
8\,B_1 \oplus 8\,B_2. 
\end{equation}
The A$_1$, B$_1$, and B$_2$ modes are infrared active, generating changes in the dipole moment along the $\hat{z}$, $\hat{x}$, 
and $\hat{y}$ directions, respectively. The A$_1$ odd-symmetry mode primarily contributes to the ferroelectric polarization in o-III phase.  

Although all four mode symmetries A$_1$, A$_2$, B$_1$, and B$_2$ are Raman active, the experimental detection of these modes in the back-scattering geometry is highly sensitive to the
orientation of the sample surface with respect to crystal lattice due to the Raman selection rules (see Supplementary Table 1). 
For instance, the B$_2$ modes can be observed only when the sample surface is parallel to the {\textbf{P}} [Supplementary Figure. 3a], whereas the A$_2$  modes can be observed only when the sample surface is perpendicular to the  {\textbf{P}} [Supplementary Figure. 3b]. 
In order to experimentally detect all the Raman-active phonons of the o-III phase, we prepared two samples having two different crystallographic surface-cutting directions. Specifically, the surface-normal ($\hat{n}$) orientations are:
(i) $\hat{n}\,\parallel \, \vec{a}$, {\textbf{P}}-in-plane and (ii)  $\hat{n} \, \bot \, \vec{a}$, {\textbf{P}}-out-of-plane. 
The exact measurement geometries are discussed in Supplementary Figure. 3.
Due to the similar lattice parameters, the $\vec{b}$ and $\vec{c}$ axes are finely twinned at the nanoscale in the polar HfO$_2$:12\% Y crystal, which means that the $\vec{b}$ and $\vec{c}$ axes are barely distinguishable in the macroscopic sample. 

Figure \ref{HfO2 12Y summary}c displays the electric polarization of the two  HfO$_2$:12\%Y crystals measured with out-of-plane electric fields. 
The samples with in- and out-of-surface plane spontaneous ferroelectric polarization are clearly identified in this measurement. 
Both samples have coercivities with similar amplitudes (approximately 3 MV cm$^{-1}$). The crystal cut to host out-of-surface plane polarization displays a larger measurable remnant polarization (on the order of 2.5 $\mu$C cm$^{-2}$) compared with the crystal cut to host an in-surface plane polarization, since the applied out-of-plane electric field is only supposed to switch the out-of-plane component of the sample polarization. It is therefore straightforward to distinguish these samples. Note that the remnant polarization value in our bulk crystals is small compared with the reported value in thin films. As discussed in Ref. \onlinecite{Xu2021}, the absence of strain from a substrate decreases the magnitude of the polar distortion and remnant polarization value. In fact, the reported ferroelectricity in HfO$_2$:Y films reveals a consistent negative correlation between remnant polarization value and sample thickness\cite{Shimizu2016,Shimura2020}.
%
%
%

Figure~\ref{HfO2 12Y summary}a,b displays the infrared reflectance and Raman scattering response of the samples with both in- and out-of-surface plane polarizations.
The infrared reflectance spectra of both samples are nearly the same, indicating that the dipole-allowed modes do not depend on the direction of electric polarization. The features observed in the reflectance spectra also nicely match with the calculated frequencies of the infrared-active phonons for the pure hafnia. 
Similar to HfO$_2$:20\%Y, the weak metallic background partially screens some of the phonon features in the infrared reflectance spectra.  
The Raman scattering spectra reveal cleaner phonon behavior.
The Raman peak positions are well resolved and are easy to track [Fig.~\ref{HfO2 12Y summary}b]. 

The crystals with the in- and out-of-surface plane ferroelectric polarization display an overall similar Raman response but with some noticeable differences. 
The relative intensities of some of the Raman-active phonon modes are slightly different due to the anisotropic character of the polarizability tensor.  
Another notable difference is that the A$_2$ and B$_1$ modes are detected only in the Raman spectrum of the {\textbf{P}} out-of-surface plane sample, whereas the B$_2$ modes are observed only in the Raman spectrum of the {\textbf{P}} in-surface plane sample. 
This occurs due to the different Raman 
selection rules for different samples. 
It is worth mentioning that the B$_1$ symmetry modes become Raman active in the {\textbf{P}} out-of-surface plane sample only because the $\vec{b}$ and $\vec{c}$ axes are mixed due to twinning. 
%
Observation of the B$_2$, A$_2$ and B$_1$ modes in our Raman measurements is completely consistent with our theoretical predictions summarized in Supplementary Table 1. 
Furthermore, our low temperature spectroscopic measurements reveal a weak temperature dependence of both the infrared reflectance and Raman scattering response in this system [Supplementary Figure. 2].

Below we highlight two peculiar phonon modes in the o-III phase.  
The first is a Hf-Hf breathing mode of B$_2$ symmetry near 252 cm$^{-1}$.
Interestingly, the displacement pattern of this Hf-Hf dimer breathing mode, shown in Fig.~\ref{HfO2 12Y summary}d, has striking similarities to the pantographic Cu-Cu dimer vibration in SrCu$_2$(BO$_3$)$_2$, which dramatically modifies the superexchange interaction by modulating the Cu-O-Cu bond angle.~\cite{Radtke2015} Of course, we do not expect the Hf-Hf dimer vibration to significantly affect the properties since hafnia is nonmagnetic. 
Moreover, a similar Hf-Hf dimer breathing mode is also observed in the experimental spectra of the orthorhombic antipolar and monoclinic phases at nearby frequencies, which implies that it is a fundamental lattice vibrational mode in the low energy phases of hafnia. 
%
%
Second, we identify a signature polar phonon mode of A$_1$ symmetry near 350 cm$^{-1}$. We call it a signature mode of the o-III phase because there is no Raman peak in all other studies phases of hafnia near the same frequency. 
The calculated displacement pattern of this A$_1$ mode, shown in Fig.~\ref{HfO2 12Y summary}e, reveals that this is the primary polar mode of the o-III phase involving polar displacements of oxygen atoms along the $\vec{c}$ lattice vector.~\cite{Lee2020, QiPRB2020, delodovici2021trilinear} 
This mode is the closest analogue to the $\Gamma_{15}^z$ polar mode that is responsible for generating scale-free ferroelectricity in hafnia and has been referred as the ``flat-phonon band" in literature~\cite{Lee2020} due to its minuscule dispersion in the momentum space.
We observe this mode in the HfO$_2$:12\%Y samples, in both in- and out-of-plane measurement configurations. However, the intensity of this peak is maximized in the sample with the out-of-plane polarization due to its relatively large polarizability matrix element for the $ac$-plane response. This mode broadens with increasing temperature [Supplementary Figure. 2], consistent with the Boltzmann sigmoid model that describes the temperature dependence of phonons. \cite{Yokosuk2015,Fan2021}  
Details of the Hf-Hf breathing and polar phonon modes in the tetragonal and o-AP phases are provided in the Supplementary Note 9.

\subsection{Fingerprinting the different phases of hafnia}

\begin{figure*}[tbh]
\centering
\includegraphics[width=6.5in]{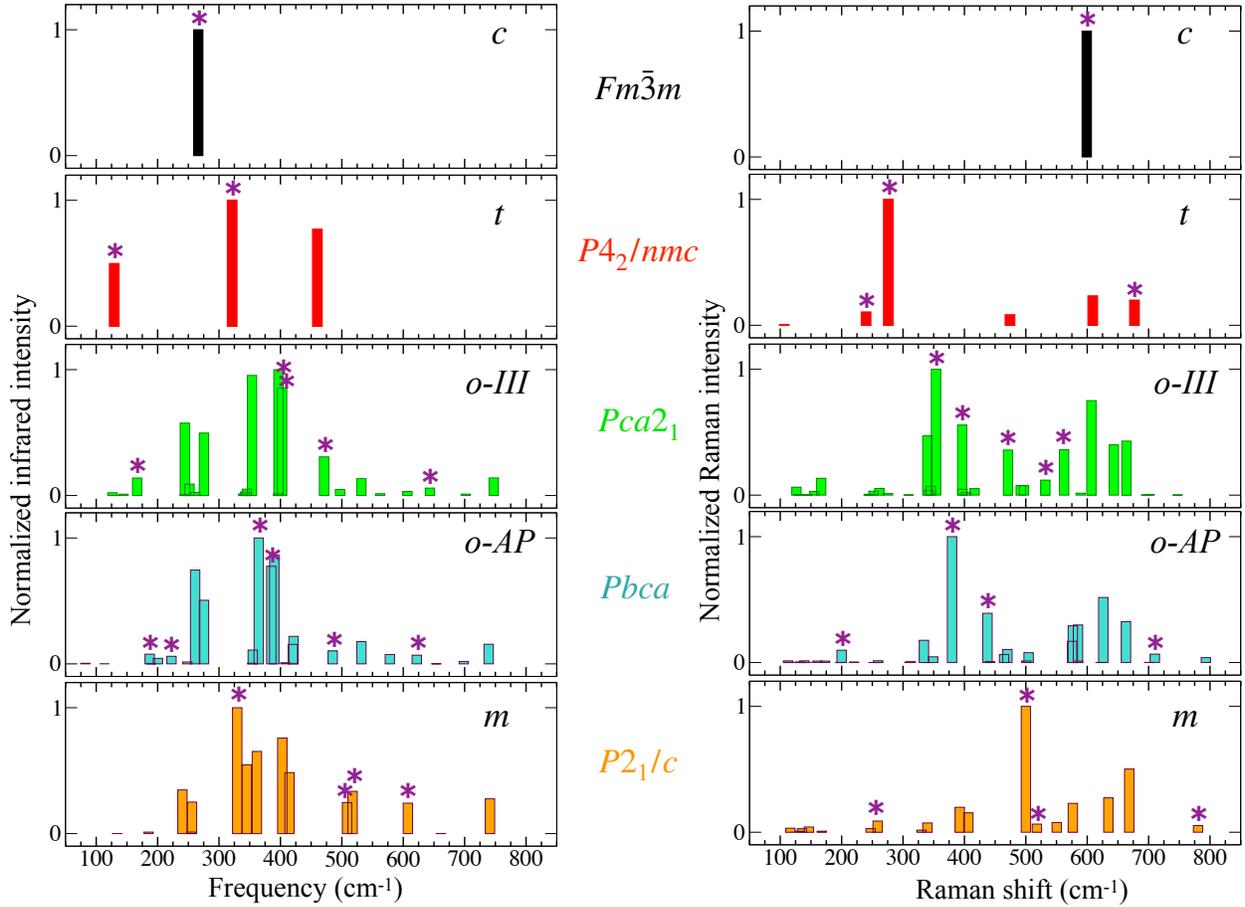}
\caption{ {\bf Spectroscopic fingerprints of different phases of HfO$_2$.} Summary of the calculated (DFT-PBEsol) infrared and Raman responses for the five studied phases of pure hafnia, {\it i.e.}, cubic (c), tetragonal (t), orthorhombic polar (o-III), orthorhombic antipolar (o-AP), and monoclinic (m). For clarity, only  TO infrared modes are provided in the calculated infrared spectra. LO mode details are available in the Supplementary Information. 
The signature modes for each phase are marked using \textcolor{purple}{$\ast$}. }
\label{Summary of all phases}
\end{figure*}

Figure \ref{Summary of all phases} summarizes the predicted vibrational properties of the known phases of hafnia. 
Comparison of the calculated peak positions and relative intensities reveals a number of distinguishing characteristics - more than enough to (i) differentiate two phases from each other or (ii) identify an unknown phase from a spectroscopic measurement. The use of vibrational spectroscopy for phase identification purposes is especially important for distinguishing between the polar and anti-polar orthorhombic phases of hafnia which are energetically and structurally very similar. 
A detailed summary of the distinguishing features 
is given in Supplementary Table 2, and as shown in Supplementary Figure. 4, these spectral fingerprints can also be used to identify mixed phases.

The cubic phase is the easiest to identify. It has only two signature modes:  T$_{1u}$ (infrared active) and T$_{2g}$ (Raman active).  
The ground state monoclinic phase is also straightforward to identify, although the number of phonons is dramatically increased due to lower crystalline symmetry. 
Distinguishing features in the infrared response include the B$_u$ symmetry modes at 330 and 518 cm$^{-1}$ as well as A$_u$ symmetry modes  at 508 and 607 cm$^{-1}$. Unique structures in the Raman scattering response include an A$_g$ symmetry peak at 500 cm$^{-1}$ and  B$_g$ symmetry modes at 518 and 780 cm$^{-1}$. The spectrum of monoclinic HfO$_2$ 
is well-documented in the literature. \cite{Belo2012, Cojocaru2019, ZhaoDVPRB2002}

Analysis is more challenging for the orthorhombic polar and antipolar phases of hafnia due to the complexity of their predicted spectral patterns along with the fact that the two phases are energetically and structurally similar. We can, however, still pinpoint several unique signatures by closely examining the calculated phonon frequencies and the relative infrared and Raman intensities in Fig. \ref{Summary of all phases}. 
In the infrared response, the A$_1$ modes at 167 and 471 cm$^{-1}$ and the B$_1$ modes at 497 and 643 cm$^{-1}$ can be used to identify the o-III phase of hafnia. 
Likewise, the B$_{3u}$ cluster near 200 cm$^{-1}$, the B$_{2u}$ modes at 385 cm$^{-1}$, and the B$_{1u}$ phonons at 390, 485, and 622 cm$^{-1}$ establish the o-AP phase of hafnia. 
In the Raman scattering response, the A$_1$ modes at 354, 396 and 471 cm$^{-1}$ and the B$_2$ modes at 532 and 562 cm$^{-1}$ are signatures of the o-III phase. They depend somewhat on the measurement geometry and different crystallographic surface-cutting directions with respect to the spontaneous ferroelectric polarization direction. 
For the o-AP phase, the Raman-active A$_g$ symmetry modes at 200, 380, and 438 cm$^{-1}$ and the B$_{2g}$ mode at 710 cm$^{-1}$ are the signature structures.

Finally, we discuss the vibrational properties of the tetragonal phase of hafnia. Although not yet grown in a  single crystal form, this phase will be readily identifiable once it becomes available. Unique signatures are predicted to include a strong  infrared-active A$_{2u}$ mode at 321 cm$^{-1}$ and an intense Raman-active A$_{1g}$ mode at 276 cm$^{-1}$. 
No infrared or Raman peaks are predicted to exist at these frequencies in any other phases of hafnia. The infrared-active E$_u$ mode at 129 cm$^{-1}$ and the Raman-active E$_g$ mode at 677 cm$^{-1}$ are additional characteristics of the tetragonal phase.

\subsection{Charge-structure-function relationships in the different phases of hafnia} 

In order to extend our understanding of structure-property relationships in this family of materials, we analyzed the calculated Born effective charges (${\bf Z^{*}}$) for each phase of HfO$_2$. 
These results are summarized in Supplementary Table 3 and are in good agreement with the previous data reported on the c, t, and m phases.~\cite{ZhaoDVPRB2002} 
We find that the ${\bf Z^{*}}$ tensor varies dramatically with the different phases of hafnia. 
In addition to the observed differences in anisotropy in the low-symmetry phases, we note that ${\bf Z^{*}}$ of Hf can acquire a value as large as +5.56\,e$^-$ in the cubic phase to relatively smaller values nearing +5\,e$^-$ in the lower symmetry phases of hafnia (lowest $Z^{*}_{\text{Hf}}$ = +4.83\,e$^-$ in the m phase).  
In each case, the ${Z^{*}_{\text{Hf}}}$ is larger than the nominal valence charge of the Hf atoms (+4\,e$^-$). Such an anomalously large ${Z^{*}_{\text{Hf}}}$, exceeding the Madelung limit, indicates a strong dynamic charge transfer from oxygen to hafnium atoms along the Hf-O bond. It further implies the mixed ionic-covalent nature of the Hf-O bonds in hafnia. Similar behavior has been observed in 1T-HfS$_2$ (${Z^{*}_{\text{Hf}}}$ = +5.3\,e$^-$).~\cite{Neal2021}

\subsection{Outlook}

Looking ahead, many of the metastable phases of HfO$_2$ discussed in this work will find  application 
in emerging chip technologies. Naturally, properties like thermal conductivity and heat capacity are key 
to effective heat management. 
Useful microscopic models of dissipation in these high $\kappa$ dielectrics will 
require information about the fundamental excitations of the lattice. 
That first principles methods can so accurately predict phonon eigenvectors and intensities bodes well for these modeling efforts. 
At the same time, this work opens the door to phononic engineering of hafnia-containing device architectures where strategies to block damaging lattice  effects can be tested 
and evaluated in advance of any measurement, significantly extending the complexity of chip modeling efforts. 

\section{Methods}

\subsection{Crystal growth and sample preparation:} 

HfO$_2$:$x$\%Y crystals were grown utilizing laser-diode-heated floating zone techniques. \cite{Xu2021} The optimal growth conditions were 95\% laser power (approximately 3000 K), atmospheric air flow of 0.1 L min$^{-1}$, and counter-rotation of the feed and seed rods at 3 and 20 rpm, respectively. Rapid cooling is one of the key factors to stabilize meta-stable phases. Therefore, growth rate of 20 mm h$^{-1}$ or a subsequent laser quench process, i.e. quickly scanning the crystal by 65\% power laser-heated zone, were performed to reach very high cooling rates. The crystal rods were oriented by Laue back-reflection x-ray methods and cut into \textit{ab/ac} and \textit{bc} plates, with typical size $\approx$2 $\times$ 2 mm$^2$. All samples were polished to reveal shiny, flat surfaces suitable for spectroscopy. A summary of the growth and processing conditions of our various samples is given in Table \ref{summary of HfO2}. 

\begin{table*}[tbh] 
\centering
\caption{\textbf{Summary of HfO$_2$ single crystals with different Y substitutions}}
\label{summary of HfO2}
\resizebox{1.0\textwidth}{!}{\begin{tabular}{|c|c|c|c|c|c|c|}
\hline
{HfO$_2$: $x$\%Y crystals}   & {20\%} & {12\%, $\textbf{P}$ in-plane} & {12\%, $\textbf{P}$ out-of-plane} & {11\%}  & {8\%}  &{0\%}     \\
\hline
Structure &  Cubic & Orthorhombic & Orthorhombic & Orthorhombic & Mixing of orthorhombic & Monoclinic \\ Phase & non-polar & polar & polar & antipolar & antipolar and monoclinic & non-polar \\

Surface cutting & Random & $\textbf{P}$ in-plane ($bc$) & $\textbf{P}$ out-of-plane ($ab$/$ac$) & Pseudo-cubic & Pseudo-cubic & Random \\

 Polarization direction &  None & $bc$-plane & $ab$- or $ac$-plane & None & None & None\\

Measurement geometry & $\textbf{k}$ $\perp$ surface & $\textbf{k}$ $\parallel$ $a$ & $\textbf{k}$ $\perp$ $a$ & $\textbf{k}$ $\perp$ surface & $\textbf{k}$ $\perp$ surface & $\textbf{k}$ $\perp$ surface \\

Growth rate & 20 mm h$^{-1}$ & 20 mm h$^{-1}$ & 20 mm h$^{-1}$ & 20 mm h$^{-1}$ & 4 mm h$^{-1}$ & 4 mm h$^{-1}$\\

Quench rate & 20 mm h$^{-1}$ & 20 mm h$^{-1}$ & 20 mm h$^{-1}$ & 20 mm h$^{-1}$ & 300 mm h$^{-1}$ & --\\
\hline
\end{tabular}} \par
\end{table*}

\subsection{Infrared and Raman scattering response:} 

We measured the infrared reflectance of a series of  Y-substituted HfO$_2$ single crystals [Table \ref{summary of HfO2}] using a series of Fourier transform spectrometers equipped with liquid helium- and nitrogen-cooled detectors. In this work, we focus on the 20 - 850 cm$^{-1}$ frequency range.  
%
Raman scattering measurements were performed in the back scattering
geometry with normal incident and unpolarized light using a LabRAM HR Evolution spectrometer. Different excitations lasers were tested and employed as appropriate to minimize the fluorescence background ($\lambda_{\text{excit}}$ = 476, 532, and 785 nm; Power $\leq$ 3.2 mW). Specifically, we used the 476 nm laser to measure the monoclinic (HfO$_2$:0\%Y),  mixed phase (HfO$_2$:8\%Y), and  cubic (HfO$_2$:20\%Y) materials. The orthorhombic antipolar (HfO$_2$:11\%Y) and polar (HfO$_2$:12\%Y) materials were measured using the 532 nm laser. 
Each scan was integrated for 60 s and averaged 5 times.  
Temperature control was achieved with an open-flow cryostat (350 - 4 K). The sample surfaces are polished to optimize the infrared reflectance and Raman scattering signal. The thickness of all samples after polishing are approximately 200 $\mu$m.

\subsection{Symmetry analysis and lattice dynamics calculations:}  

First principles density functional theory (DFT) calculations were performed using the projector augmented wave method as implemented in the Vienna Ab initio Simulation Package (VASP).~\cite{Kresse96a, Kresse96b, KressePAW} The exchange-correlation functional was computed using the generalized-gradient approximation as parameterized by Perdew-Burke-Ernzerhof for solids (PBEsol).~\cite{PBEsol} 600 eV was used as the kinetic energy cutoff for plane waves. The energy convergence criterion for  self-consistent DFT calculations was set at $10^{-7}$ eV and force convergence criterion for relaxation of atomic coordinates was set at $10^{-3}$ eV \AA$^{-1}$. The reciprocal space of the cubic (c), tetragonal (t), orthorhombic polar (o-III), orthorhombic antipolar (o-AP), and monoclinic (m) phases was sampled using a Monkhorst-pack $k$-mesh~\cite{MP1976} of size 8$\times$8$\times$8,
12$\times$12$\times$8,
8$\times$8$\times$8,  8$\times$8$\times$4, and 
8$\times$8$\times$6, respectively. 
The phonon dispersions of the c, t, o-III, o-AP, and m phases were calculated using supercells of size 4$\times$4$\times$4, 4$\times$4$\times$2, 2$\times$2$\times$2, 2$\times$1$\times$2, and 2$\times$2$\times$2, respectively. 
The non-analytical term correction was employed using the Gonze scheme~\cite{GonzePRB1994, GonzePRB1997} as implemented in the {\sc phonopy} package.~\cite{phonopy} 
A commensurate $k$-grid was used for phonon calculations. 
The zone-center phonon eigenvectors were used to compute the theoretical infrared and Raman spectra using the methods described in Ref.~\cite{Skelton2017} 
An appropriate averaging of the Raman-activity tensor was done to simulate the Raman spectrum of each studies phases of hafnia. 
The Bilbao Crystallographic Server,~\cite{bilbao2003} {\sc phonon} website: http://henriquemiranda.github.io/phononwebsite/phonon.html, and {\sc phonopy} package~\cite{phonopy} were utilized to determine the Raman selection rules and symmetry of the phonon modes. 

\section{\bf Data availability} 

Relevant data are available upon request from the corresponding authors.

\section{Acknowledgements}  
S.F., K.P., and J.L.M. appreciate support from the National Science Foundation (DMR-2129904 and DMR-1629079) and the Materials Research Fund at the University of Tennessee. 
X.X. and S.W.C. were supported by the center for Quantum Materials Synthesis (cQMS), funded by the Gordon and Betty Moore Foundation’s EPiQS initiative through grant GBMF6402, and by Rutgers University.
S.S., Y.Q., and K.M.R. acknowledge funding from Office of Naval Research (ONR) grant N00014-21-1-2107. 
D.V. acknowledges support from National Science Foundation grant DMR-1954856.
First-principles calculations were performed using the computational resources provided by the Rutgers University Parallel Computing clusters. This work was supported by the U.S. Department of Energy (DOE), Office of Science, Basic Energy Sciences under award
DE-SC0020353 (S. Singh).

\section{Competing Interests}

The authors declare no competing interests.

\section{Author contributions} 
Shiyu Fan and Sobhit Singh contributed equally to this work. This project was conceived by J.L.M. and S.W.C. The single crystals were grown and polarization loop measured by X.X. with advice from S.W.C. 
The spectroscopic work was performed by S.F. and K.P. with advice from J.L.M.  The first principles DFT calculations were carried out by S.S. with advice from D.V. and K.M.R. Data analysis was completed by S.F., S.S., K.P., Y.Q., D.V., K.M.R., and J.L.M. All authors discussed the findings. The manuscript was written by S.F., S.S., X.X., K.P., K.M.R., and J.L.M. All authors read  and commented on the manuscript.

\section{Corresponding Authors} Direct correspondence to Janice L. Musfeldt (musfeldt@utk.edu) and Karin M. Rabe (rabe@physics.rutgers.edu).

\bibliography{hafnia_ref.bib}

\end{document}